Orbit symmetry breaking in MXene implements enhanced soft bioelectronic implants

Yizhang Wu[1+], Yuan Li[2+], Yihan Liu[1+], Dashuai Zhu[3], Sicheng Xing[1], Noah Lambert[1], Hannah Weisbecker[1], Siyuan Liu[1], Brayden Davis[1], Lin Zhang[1], Meixiang Wang[4], Gongkai Yuan[1], Chris Zhoufan You[5], Anran Zhang[1], Cate Duncan[1], Wanrong Xie[1], Yihang Wang[1], Yong Wang[6], Sreya Kanamurlapudi[2], Garcia-Guzman Evert[1], Arjun Putcha[1], Michael D. Dickey[4], Ke Huang[3]*, Wubin Bai[1]*

[1]Department of Applied Physical Sciences, University of North Carolina, Chapel Hill, NC 27514, USA

[2]Joint Department of Biomedical Engineering, North Carolina State University; University of North Carolina at Chapel Hill, Raleigh, North Carolina, 27607, USA

[3]Department of Molecular Biomedical Sciences, North Carolina State University, Raleigh, North Carolina, 27607, USA

[4]Department of Chemical and Biomolecular Engineering, North Carolina State University, Raleigh, North Carolina, 27606, USA

[5]Chapel Hill High School, Chapel Hill, North Carolina, 27516, USA

[6]Department of Physics, Nanjing University, Nanjing, Jiangsu, 210000, China

[+]Contribution equally

*Corresponding author: khuang7@ncsu.edu & wbai@unc.edu

**Summarize in one sentence:** Symmetry engineering in MXene features the improved bioelectronic-tissue impedance, facilitating the bioelectronic sensing interfaces.

## Abstract

Bioelectronic implants with soft mechanics, biocompatibility, and excellent electrical performance enable biomedical implants to record electrophysiological signals and execute interventions within internal organs, promising to revolutionize the diagnosing, monitoring, and treatment of various pathological conditions. However, challenges remain in improving excessive impedance at the bioelectronic-tissue interface and thus the efficacy of electrophysiological signaling and intervention. Here, we devise orbit symmetry breaking in MXene (a low-cost scalability, biocompatible, and conductive 2D layered material, that we refer to as OBXene), that exhibits low bioelectronic-tissue impedance, originating from the out-of-plane charge transfer. Furthermore, the Schottky-induced piezoelectricity stemming from the asymmetric orbital configuration of OBXene facilitates interlayered charge transport in the device. In this study, we report an OBXene-based cardiac patch applied on the left ventricular epicardium of both rodent and porcine models to enable spatiotemporal epicardium mapping and pacing, while coupling the wireless and battery-free operation for long-term real-time recording and closed-loop stimulation.

**Introduction**

Soft bioelectronic implants, melding flexible, conductive materials, enable conformal interfacing capabilities, facilitating real-time, precise electrophysiological mapping and interventions(*1*). It promises to fulfill unmet clinical needs by entitling long-term implants to be safer, more intelligent, more functional, and easier to deploy, thereby improving the patient quality of life(*2, 3*). Bioelectronic-tissue impedance is a pivotal factor in the development and functionality of these implants, critically influencing the design, operational efficiency, and long-term stability of implantable devices within the human body(*4, 5*).

Monolayer MXene flakes, featuring solution processability(*6*), enable terrific intra-layer conductivity(*7*), assigned to transition metal atoms in MXenes hybridized with the carbon atoms, leading to an inner spread of electronic states, developed as epidermal soft electrode "MXtrode"(*8*). However, MXenes suffer from a low out-of-plane transport of charge carriers, thus often compromising the interfacial impedance at the bioelectronic-tissue interface, especially in in-body organs(*9*). Such impedance challenges can negatively impact electrophysiological signal fidelity and the effectiveness of interventions, thereby posing constraints on the integration of MXenes in soft bioelectronic implant applications(*10*).

Here, to lower interfacial impedance in MXenes, and widely in soft bioelectronic implants, we induce orbit symmetry breaking of *p-d* hybridized orbitals in MXene, which we refer to as OBXene, by subjecting MXene to a mild facial oxidation process. This perturbation, while slightly compromising the inherent in-plane electrical conductivity of MXenes, remarkably gives rise to out-of-plane charge transfer capabilities. Furthermore, we observe the Schottky-induced piezoelectric effect in OBXene, attributable to asymmetric orbital configurations, which enhances out-of-plane charge transport and leads to ultra-high sensitivity to mechanical deformation featuring as fillers in elastic composites.

To cast fundamental insights into practical bioelectronic implants, we designed an OBXene-based cardiac patch to enable real-time physiologic monitoring of the heart in both small animals and clinically relevant large animals in various myocardial infarctions stages. This patch features (i) low interfacial impedance at the bioelectronic-tissue interface, (ii) high conductivity, (iii) long-term biocompatibility, (iv) versatile utilization in sensor formation by combining with ionogels, and (v) low-cost scalability. This patch enables high-fidelity spatiotemporal mapping of physiological activities and multimodal therapeutic capabilities. *In-vivo* studies using a porcine model demonstrate its wireless, battery-free operation for prolonged implantation, providing real-time diagnostics and closed-loop electric stimulation. The optimization of interfacial impedance and long-term biocompatibility of OBXene traces also approach them as excellent coatings on the electrode surface of existing commercially implantable electronics including cardiac pacemakers, neurostimulators, deep brain stimulators, and others, for improved performance.

**Results**

**Flexible and scalable cardiac patch based on OBXene traces**

Cardiovascular diseases (CVDs) are a major global health concern, generating an annual expenditure of 138.33 billion USD in medical treatment(*11*). CVDs, particularly myocardial infarctions (MI), are still the leading cause of death, accounting for approximately 31 % of annual global deaths and 17.9 million fatalities(*12*). In response to this challenge, we designed an integrated, multimodal OBXene cardiac patch, highlighting its functionalities, mechanical flexibility, and material composition, which could be deployed to left ventricle (LV) epicardium to offer multimodal sensing and electrotherapy to the infarcted myocardiocytes from occlusion of the left anterior descending branch (**Fig.1A**). The high solution processability, enabled by the hydrophilic terminating functional groups of MXene, allows its uniform coating onto a stretchable elastomeric film based on styrene-block-(ethylene-co-butylene)-block-styrene (SEBS) and chitosan (a biocompatible polymer approved by the U.S. Food and Drug Administration, FDA). Then, a beam of near-infrared (NIR) laser effectively patterns the coated MXene layer by locally ablating it without observable damage to the SEBS layer underneath (Fig.S1). After laser patterning, a crucial step involves a brief exposure to oxygen plasma to achieve a simple oxidation process, which is essential for deriving anatase $TiO_2$ on the MXene and introducing breakage of the orbital symmetry in the MXene. An exploded view of the OBXene patch highlights its multilayer integration with a variety of OBXene-based sensors, stimulators, and serpentine interconnects (**Fig.1B**). A thin layer of biocompatible parylene C (thickness 5 um) coated via chemical vapor deposition fully encapsulates the OBXene traces (thickness 3 um) with selected regions exposed for targeted electrical contact with tissues. The OBXene patch consists of an electrocardiography (ECG) mapping array, a pressure mapping array, a strain sensor, a temperature sensor, a heater, and an electrical stimulator (**Fig.1C**). Collectively, these components not only monitor the electrical, thermal, and contractile physiology of local cardiomyocytes in real time but also perform electro and thermal therapies with high spatiotemporal precision.

Notably, the pressure mapping array features an OBXene-filled ionogel (poly(acrylic acid-co-acrylamide, P(AA-co-AAm)) sandwiched by two OBXene pads to offer ultrahigh sensitivity by combining the capacitive, resistive, and piezoelectric responsivities to deformations of neighboring myocardium. The surface electronegativity of MXene, resulting from its chemical exfoliation, facilitates its self-assembly with electropositive chitosan and promotes coating adhesion with SEBS to ensure integration stability and robustness. The integrated OBXene patch also exhibits high mechanical flexibility and softness as it undergoes bending and crumpling processes (**Fig.1D**). OBXene flakes approximately 1.75 nm thick and up to micrometers in lateral size (Fig.S2a) show observable distinctions in their morphologies before and after oxidation. The TEM pattern (**Fig.1E,** S2b&c) presents the fusiform anatase $TiO_2$(*13*) anchored at the border of the OBXene flakes, which differs from the smooth border in pristine MXene. Derived anatase $TiO_2$ produces a charge transfer path different from direct in-plane electronic transport in pristine MXene, underscored by the comparative surface photovoltage responses (SPV, Fig.S2d&e). The oxidation process plays a pivotal role in regulating charge behavior within OBXene. The cross-sectional view (**Fig.1F**), while revealing the layered configuration, further highlights the incorporation of oxides on the edge of the MXene's outer surface. The derived oxides arise from subjecting patterned MXene traces to oxygen plasma irradiation during the fabrication of the OBXene patch. The OBXene-based cardiac patch enables soft implantable mechanics and seamlessly integrates multimodal sensing and therapy modules. Sustainability is essential for developing and manufacturing biomaterials, especially at scale(*14*). This cardiac patch enables integrated multimodal sensing from a central platform, minimizing material redundancy and reducing potential waste.

**Orbit symmetry breaking in MXene facilitates soft bioelectronic implants**

OBXene activates out-of-plane charge transfer pathways *via* derived Ti-O bonds, consequently resulting in an ideal impedance at the bioelectronic-tissue interface (**Fig.2A**). Fourier transforms (Fig.S3a) of the extended X-ray absorption fine structure (EXAFS) and corresponding Morlet wavelet transformation ($k^2$-weighted, **Fig.2B** and Fig.S3b) in R-space elucidate the derived bonding configuration in OBXene, featuring Ti-O bonding, which is distinguished from various titanium species (e.g., compounds, elements, and oxides)(*15*). Meanwhile, OBXene shows a dipole transition originating from the 1 s orbital of the Ti atom to the *p*-component of the *p-d* hybridized orbital (Fig.S3c). Density functional theory (DFT) calculations illustrate derived oxygen atoms in OBXene that possess iconicity (Fig.S3d), serving as a "bridge" facilitating charge transfer capability. It is noteworthy that the orbital symmetry arising from *p-d* hybridization (Fig.S3e&f) is disrupted due to the significant enhancement of the state density of oxygen projected in the $P_z$ orbital (**Fig.2C**)(*16*). Diffraction and spectroscopic investigations (Fig.S4) provide insights into the internal structure of OBXene, focusing on bonding, chemical composition, and charge migration. Specifically, it is noted that (i) Orbit symmetry breaking in OBXene effectively regulates the charge transfer pathways, facilitating out-of-plane migration. (ii) OBXene indicates nonuniform out-of-plane vibrations attributed to the distinct terminal groups, which should be duly considered in subsequent theoretical calculations. (iii) Charges migrate from semiconductor anatase $TiO_2$ to metallic MXene, thereby indicating the presence of the interfacial Schottky barrier in OBXene. The Schottky interface may manifest piezoelectric effects in centrosymmetric anatase $TiO_2$ induced by interface symmetry breaking(*17*).

Low bioelectronic-tissue impedance is crucial for electronic implants, as it plays a pivotal role in mitigating electrical noise and ensuring high signal-to-noise ratios (SNRs) during the recording process. **Fig.2D** and Fig.S5 show a comparative analysis of various bioelectrode materials during their immersion in 1x phosphate-buffered saline (PBS) aqueous solution, including OBXene, MXene (pristine monolayer), Au (precious metal), graphene oxide (GO, porous oxide), and PEDOT:PSS (conductive polymer), on their galvanostatic charge-discharge (GCD), cyclic voltammetry (CV), and electrochemical impedance spectroscopy (EIS) to assess their corresponding interfacial impedance and charge-storage capacity (CSC). GCD curves (Fig.S5a) exhibit a subtle nonlinearity indicative of pseudocapacitive behaviors within OBXene. In the case of CVs for OBXene, we observed broadened peaks with minimal separation in peak positions during charge/discharge cycles. In contrast, CVs for GO present a typical rectangular shape, consistent with the electrical double-layer capacitor (EDLC) mechanism, where instantaneous charge separation occurs without diffusion limitations. It is noteworthy that both pseudocapacitors and EDLCs involve faradaic reactions in charge storage. However, the pseudocapacitive behavior in OBXene primarily arises from surface redox pseudocapacitance, driven by fast ion intercalation and changes in the oxidation state, particularly associated with the derived anatase $TiO_2$. Conversely, the charge storage process in GO originates from the electrosorption of ions by porous carbon, driven by electrostatic forces(*18*). Pseudocapacitive behavior in OBXene boasts a wider electrochemical stability window (-0.5 to 0.5 V) compared to EDLC-based GO, with minimal distortion, particularly in the cathodic region. This extended potential window holds promise for electrotherapy applications. The OBXene electrode exhibits higher current densities and charge storage capacity (Fig.S5b&c) at the electrode surface compared to the other three types of electrodes, aligning with improved charge injection capacity (CIC). The presence of edge effects arising from the derived $TiO_2$ predominantly influences the charge-carrying capacity of the border region, resulting in a more centralized linear diffusion process for ions(*19*). This increase in CSC and CIC equips the OBXene electrodes to safely administer charge injection at biologically relevant levels, such as in electrical pacing to synchronize electrophysiological signals in myocardial tissues.

EIS measurements (Fig.S5d) highlight the reaction kinetics, with a specific focus on the charge transfer resistance ($R_{ct}$). This resistance parameter features the diameter of the semicircle arc. The measured $R_{ct}$ values for the electrodes with the same geometry based on OBXene, MXene, Au, GO, and PEDOT:PSS are 144 Ω, 115 Ω, 33 Ω, 2775 Ω, and 924 Ω, respectively, indicating low charge transfer resistance coupled with fast ion diffusion/transport kinetic of OBXene, which facilitates high SNR data transmission, lowers power consumption, and enables wireless power transfer within bioelectronic implants. OBXene electrode consistently displayed the lowest interfacial impedance within the physiologically relevant frequency range of $10^{-2}$ Hz to $10^4$ Hz over other conventional implantable electrodes (**Fig.2D** and Fig.S5e&f), encompassing electrophysiological relevance of electroactive myocardium tissues (0.02 Hz to 100 Hz). Hence, the enabled out-of-plane charge transfer led by orbital symmetry breaking in OBXene significantly facilitates charge transport between the bioelectronic surface and the tissue epidermis.

OBXene holds great promise as a far-reaching bioelectronic interface with broad scenarios for application, not only attributed to its remarkable electrochemical impedance and also a cohort of other properties, such as scalability, visibility, conductivity, and biocompatibility, that are equally essential in biomedical applications.

**Fig.2E-H** provide a systematic investigation of biocompatibility in the OBXene electrode in comparison with several conventional counterparts, including MXene, Au, GO, and PEDOT:PSS. Low cytotoxicity is essential for electrode biosafety. To evaluate the cytotoxicity of the OBXene electrode, we cocultured OBXene with neonatal rat cardiomyocytes (NRCM; detailed procedures appear in the Methods section) for 48 hours. The LIVE/DEAD analysis results indicated that there were no significant NRCM viability differences between the group immersed with the OBXene electrode and the control group with no electrodes, while the groups with MXene, PEDOT:PSS, Au, and GO all revealed slightly decreased cell viability compared to the control group (**Fig.2E, I**). Another key determinant of cytotoxicity is the cell metabolic activity of NRCMs. Our result of the cell counting kit-8 (CCK8) assay revealed that the OBXene electrode demonstrated metabolic activity consistent with that of the control group and significantly higher than those of all other experimental groups (**Fig.2J**). Then, we replicated the experiments by using H9C2 cardiomyocytes to further validate the low cytotoxicity of OBXene observed in NRCM-based assays. Then, we used the LIVE/DEAD and CCK8 analyses, and the result indicated that contact with the OBXene electrode for 48 hours stimulated a higher viability and metabolic activity in H9C2 cells when compared to other types of electrodes (Fig.S6). To gain a comprehensive understanding of the biocompatibility of OBXene electrodes *in vivo*, we implanted the electrodes subcutaneously in the abdominal area of Sprague Dawley rats for periods of 7 and 14 days to evaluate the short-term and long-term immune responses. The use of immunohistochemistry (IHC) on the rats, targeting CD68 (macrophage) and CD11b (neutrophil) markers at 7 days, facilitated the measurement of the innate immune reaction in the skin tissue. Further, characterization using the CD3 (T cell) marker on rat skin tissue at 14 days provided insights into relatively longer-term immune reaction characterized by T-cell infiltration (**Fig.2F**). The results showed that the OBXene electrode induced less inflammation compared to all the other electrodes, as evidenced by significantly fewer CD68+, CD3+, and CD11b+ cells (**Fig.2K**). Masson's trichrome (**Fig.2G** and Fig.S7) and hematoxylin and eosin (H&E) histology (**Fig.2H**) visualized the surrounding skin tissue fibrosis and inflammatory cell infiltration. These results demonstrated that the OBXene electrode exhibited significantly less fibrotic tissue

formation compared to the MXene, PEDOT:PSS, Au, and GO electrodes (**Fig.2L**). Overall, the OBXene electrode exhibited desirable biocompatibility, as illustrated by minimal cytotoxicity *in vitro*, low inflammation *in vivo*, and thinner fibrotic tissue formation compared to the typical electrodes used in implantable electronics.

Operational stability is another paramount attribute of implants in their practical utility. Fig.S8 demonstrates that the OBXene electrode can maintain its electrical conductivity within a 10 % deviation for more than 30 days of immersion in PBS solution, outperforming the other three types of bioelectrodes. We attribute this stable longevity to effective bonding via chitosan cross-linking at the interface between OBXene and SEBS, resulting in the formation of dense and stable OBXene electrodes. Such system stability not only obviates the need for frequent replacements or adjustments but also ensures the dependable and uninterrupted delivery of precise therapeutic interventions, including electrical stimulation and thermal ablation.

Optimal imaging visibility is vital for electronic implants because it ensures accurate implantation and convenient follow-up device inspection, safeguarding both therapeutic efficacy and clinical safety over the long term. The presence of unpaired electrons on Ti atoms (projected on the $P_z$ orbit) in OBXene imparts a paramagnetic character, potentially influencing local magnetic fields and enhancing proton relaxation rates. **Fig.2M** and Fig.S9 show the *in vitro* $T_1$- and $T_2$-weighted MRI characterization of OBXene, conducted under physiologically relevant neutral pH conditions resembling cardiac biofluid, using an ultrahigh field (7.0 T) MRI scanner. The results demonstrate that OBXene in isolation can possess intrinsic MRI contrast capabilities. Whereas MXene, Au and PEDOT:PSS remained imperceptible in imaging. This visualization remains undetectable in pristine MXene(*8*), highlighting the benefits of symmetry breaking in implantable soft bioelectronic systems.

As summarized in **Fig.2N**, OBXene shows optimized electrochemical impedance while exhibiting comprehensive performance serving as a soft bioelectronic interface. Complemented by its outstanding biocompatibility, visibility, scalability relative to conventional electrode materials for implantable electronics, OBXene promises to be a prominent candidate in the realm of bioelectronic materials.

## *In vivo* high-fidelity recording and effective electrotherapy

To demonstrate the advantageous properties of OBXene as a bioelectronic implant, systematic studies using a rat MI model on individual functional modules constituting the OBXene patch are provided in **Figs.3-5**.

Electrocardiography (ECG) serves as a valuable diagnostic tool in an MI model, providing crucial insights into the pathophysiological aspects of blood flow obstruction. High-fidelity recording in OBXene-based electrodes, attributable to low bioelectronic-tissue impedance, can enable real-time monitoring of cardiac rhythm and conduction abnormalities(*20*), and augment the assessment of therapeutic interventions and prognostic outcomes. This device is engaged with a live rat heart through cardiothoracic implantation featuring two OBXene-based serpentine electrodes, one for capturing an averaged ECG (AECG) over the interfaced area of the rat heart and the other for supplying electrical stimulation (**Fig.3A**). **Fig.3B** shows a representative recording of the AECG from the OBXene electrodes, highlighting its sensing stability and fidelity. Electrical stimulation offered by the OBXene electrode with various stimulation parameters (unipolar stimulation, with amplitude ranging from 1 V to 5 V and frequency ranging from 1 Hz to 10 Hz.) can effectively restore the cardiac waveforms from a distorted, weak infarcted state to a strong, regular state (**Fig.3C** and Fig.S10a-c). We employed a methodical approach, placing a 3 x 3 electrode array centered on the left ventricle and collecting ECG from multiple directions to ensure comprehensive signal assessment. Fig.S10d&e reveal that OBXene electrodes consistently outperform gold electrodes in terms of signal-to-noise ratio (SNR) and signal intensity, validating the superior signal quality obtained through OBXene, attributable to the reduced bioelectronic-tissue interfacial impedance facilitated by the orbital symmetry breaking.

**Fig.3D** further highlights the regulated pacing facilitated by electrotherapy, comparing the cardiac waveforms in the frequency spectrogram. Specifically, the ECG measurements from the OBXene patch show that the MI model exhibits a prolonged R-wave (represents the early ventricular depolarization) duration compared to normal conditions, signifying conduction disturbances. Detailed analysis reveals both a decrease in R-wave amplitude and an elevation-displacement in the ST-segment (represents the interval between ventricular depolarization and repolarization) of the infarcted heart, indicating acute myocardial ischemia(*21*). Furthermore, the QRS complex (represents ventricular depolarization), spanning 10 to 100 Hz, shows a broadening in the MI cardiac waveforms, suggesting slow ventricular depolarization and conduction system dysfunction. Electrical pacing, particularly with a square wave of 1 V amplitude and 10 Hz frequency, corrects ST-segment displacement, indicating effective relief from acute ischemia. However, the remaining abnormal QT interval, reflecting the total duration of ventricular depolarization and repolarization, may hint at a certain risk of ventricular arrhythmias requiring additional therapeutic intervention.

The OBXene-based multichannel ECG electrode (MECG, **Fig.3E**) provides a continuous map of the electrical activity of the epicardium with high spatiotemporal resolution over a 3 mm x 8 mm area of rat myocardium. This facilitates the identification and localization of areas affected by ischemia, infarction, or specific arrhythmogenic foci. As shown in **Fig.3F**, MECG recordings can effectively reveal distinct electrophysiological signatures corresponding to various areas of a live rat heart. For example, for the MI model (**Fig.3G**), the multichannel waveforms exhibit evident ST-segment displacement, indicating varying degrees of myocardial ischemia compared to sinus rhythm. Notably, recordings of Channel 2 show the disappearance of the ST segment, pinpointing nearby subendocardial ischemia. The consistent prolongation of the R-wave peak time observed signifies abnormal LV depolarization in the MI model, resulting in impaired electrical conduction from the endocardium to the epicardium(*21*).

OBXene exhibits enhanced CIC, enabling precise charge injection at biologically relevant levels with relatively low thermal dissipation. Accurate and targeted electrical pacing plays a pivotal role in MI therapy by facilitating the resynchronization of ventricular contractions, thus contributing to myocardial recovery and the post-MI remodeling process(*22, 23*). For instance, patients with broader QRS complexes, which indicate a left bundle branch block or other intraventricular conduction delays, have demonstrated enhanced therapeutic benefits from electrical pacing interventions(*24*). **Fig.3H** shows a design that integrates MECG and AECG electrodes to achieve cardiac electrotherapy orchestrated by real-time ECG signal feedback. Here, the AECG electrode with a large-area coverage captures ECG signals, while one of the MECG electrodes serves as the cathode, and the remaining electrode functions as the anode to establish a bipolar configuration for achieving cardiac stimulation. **Fig.3I** presents the performance of concurrent ECG detection and electrical treatment by plotting the corresponding frequency-domain spectrum (FDS) of the measured ECG signals with various stimulation parameters. In contrast with the time-domain spectrum (TDS, Fig.S11), the FDS further highlights the electrotherapeutic outcomes by filtering and decomposing, to a certain extent, waveform components (P, Q, R, S, T) for targeted analysis of cardiac physiology(*25*). The FDS pattern corresponding to the MI model reveals significantly narrowed peaks in the QRS complex region, indicating a reduction in the occurrence of QRS characteristic peaks. This reduction can be attributed to an atrioventricular block caused by myocardial infarction, resulting in interrupted pulse transmission between the atria and ventricles. Atrial and ventricular excitation consequently become desynchronized, leading to a decreased heart rate. Ventricular and atrial systolic resynchronization experiences a notable improvement upon electrotherapy. Furthermore, the inverted T waves in the QT intervals of a rat MI model (**Fig.3I**) show no momentous change with the presence of electrical stimulation, suggesting that the rapid repolarization of contractile cells remains insensitive to electrical stimulation. Ventricular arrhythmias resulting from myocardial ischemia persist, aligning with the findings in the TDS. The findings illustrate the capability of OBXene electrodes to facilitate both high-quality cardiac recording and mapping, enabling detailed analysis, along with the efficient stimulation of the cardiac epicardium through effective charge transfer.

### *In vivo* multimodal sensing and thermotherapy

Cardiac tissue temperature may rise by 0.5°C to 2°C in the infarcted area, compared to non-affected tissue in MI patients(*26*). This increase is due to cellular and biochemical reactions, including inflammation and heightened metabolic activity in response to tissue damage(*27*). These temperature changes are localized, often undetectable by standard body temperature measurements, necessitating advanced medical sensors for accurate detection and treatment guidance. Additionally, thermotherapy that modulates the temperature of the epicardial surface can effectively alter the QT interval, while influences the speed and duration of cellular depolarization and repolarization(*28*). Thus, monitoring and modulation of epicardial temperature can compensate for electrotherapy, especially in the MI model.

OBXene-based cardiac patch enables the integration of temperature sensor and heater (**Fig.4A**). **Fig.4B** shows the synergistic performance between the thermal sensor and heater, displaying an ideal linear regression across various power inputs to the heater, with resulting local temperature modulation spanning from room temperature to myocardial temperatures of clinical relevance(*29*). Fig.S12a demonstrates the sensing stability of the thermal sensor for multiple cycles of thermal fluctuations achieved by the nearby thermal heater (input power, 5 W); additional characterizations with input power ranging from 5 to 10 W appear. Furthermore, Fig.S12b-d show the heating stability, reversibility, and responsivity of the serpentine-shaped heater at various power inputs (from 1 W to 5 W). Fig.S12e&f demonstrate the efficacy of the OBXene-based heater, characterized by its high-efficiency thermal response. This is evidenced by the device's capability to achieve a stable and rapid temperature increase within a timeframe of 600 seconds. Fig.S12g&h illustrate the combination of ECG electrode-heater deployed on MI myocardium, featuring an enhancement in ventricular depolarization rates, suggesting improved electrical-pulse propagation through the ventricles following thermotherapy.

**Fig.4D** shows images of a rat epicardial surface before, during, and after thermal ablation offered by a conformally attached OBXene device coupling a thermal sensor and heater, demonstrating its capability in both targeted thermal necrosis and real-time monitoring to ensure precise thermotherapy (**Fig.4C**). Higher core-body temperature is associated with enhanced cardiac output, increased vasodilation in situ and the peripheral infarct region(*30, 31*). These thermotherapeutic benefits may not directly treat myocardial infarction but can prevent further complications and alleviate the symptoms for patients.

Monitoring myocardial mechanics, especially the left ventricular ejection fraction (LVEF), is essential in post-MI cardiac assessment(*32*). A continuous evaluation of both systolic and diastolic mechanics can provide deep insights into both sudden impairment of cardiac function and its recovery post-recent cardiac arrest(*33*). **Fig.4E** and Fig.S13a show the OBXene-based strain sensor, highlighting its mechanical flexibility and conformal attachment with curvilinear and successive pacing of the myocardium. **Fig.4F** illustrates the measured calibration curve of the sensor, depicting a linear relationship between the strain and the resistive sensing signal with little attenuation after multiple cycles of repetitive loading and unloading (Fig.S13b&c). Notably, the strain sensor exhibits ultrahigh sensitivity to pressure, with a detection threshold as low as 371 Pa. As indicated in Fig.S13d, the sensor demonstrates a positive linear correlation between $\Delta R/R_0$ and pressure. Moreover, Fig.S13e shows a fast response time (< 100 ms) without hysteresis, enabling the strain sensor to capture myocardial contractions in real time. The strain sensor incorporates a Wheatstone bridge (Fig.S13f) with temperature compensation to mitigate temperature-induced interference on the strain sensor during its *in vivo* operation. Fig.S13g-j exhibit the deployment of OBXene-based Wheatstone bridge on rat myocardium, showcases its capability to capture physiological signals with enhanced baseline stability.

**Fig.4G** illustrates the deployment of an OBXene-based strain sensor on a living rat heart in conjunction with a multichannel electrode for simultaneous electrotherapy. **Fig.4H** shows the measured contractile change in the beating heart under various conditions ranging from the normal state to the normal state with electrotherapy ($Sq_{10}^1$), the MI state, and the MI state with electrotherapy. MI states exhibit compromised cardiac contraction, with a slower beating rate and weaker contractility in comparison to those of the normal state. Electrical pacing interventions enhance the rhythm and contractile performance of infarcted hearts. The strain sensor provides a feedback assessment of cardiac contractility, highlighting its high promise to guide and optimize electrotherapy in real-time and potentially enable automation.

**OBXene-filled ionogels with piezoelectric properties for pressure mapping**

Visualizing myocardial mechanics plays a pivotal role in not only the precise identification and characterization of impaired myocardial segments but also in gaining a fundamental understanding of MI pathology with high spatiotemporal resolution mapping. Our design uses an active matrix based on OBXene electrodes integrated with a pressure-responsive ionogel to achieve this goal. The ionogel relies on a randomly copolymerized network of acrylamide (PAA) and acrylic acid in 1-ethyl-3-methylimidazolium ethyl sulfate (PAAm), solvated in an ionic liquid (1-ethyl-3-methylimidazolium ethyl sulfate, or EMIES), and has demonstrated ultratough and stretchable mechanics(*34*). Introducing OBXene as a filler into the ionogel further optimizes its elasticity to physiological relevance and enhances its electrical response to external pressure, as demonstrated in **Fig.5**.

A 124-pixel mapping array based on OBXene-filled ionogel is situated between two OBXene contact pads supported on flexible SEBS substrates (**Fig.5A**). This design offers conformable contact even under deformations, meeting the required mechanical requirements for engagement with myocardial tissue. Here, the electronegative OBXene fillers are added to ionogels, as illustrated in **Fig.5B**. In the absence of the OBXene fillers, the ionogels are known to have a large modulus due to the presence of polymer-rich nanodomains that phase separate from the otherwise solvated gel(*35*). Interestingly, the addition of OBXene to the ionogels prevents the formation of these polymer-rich domains by presumably disrupting hydrogen bonding between the polymer chains. Consequently, the composite ionogels are very soft. Fig.S14 demonstrates the stable stretchability of the OBXene-filled ionogel, with high fracture strain (~ 944 %). The incorporation of OBXene (2 wt.%) into the ionogel lowers Young's modulus compared with that of the pristine ionogel, thus enhancing its conformability when interfacing with internal organs, including the lung, brain, and heart, as shown in **Fig.5C**. In addition, increasing the filling ratios of OBXene (e.g., from 0 wt.% to 10 wt.%) in the ionogel can effectively decrease the tensile strength (**Fig.5D**), thus offering a tuning window to optimize the system mechanics to match those of targeted organs in contact (**Fig.5E**) for enhanced biocompatibility.

OBXene can further amplify the pressure sensitivity incorporated into the ionogel network (**Fig.5F-K**). Briefly, OBXene capitalizes on the derived anatase $TiO_2$ coupled with metallic MXene, forming a Schottky junction. At the Schottky interface, a built-in electric field emerges that breaks the inversion symmetry and induces polar structures, resulting in emergent piezoelectricity(*17*). The piezoelectric effects, acting as strain accommodation, improve interlayered electron transport by preventing significant increases in interfacial resistance caused by structural deformation(*36*). Figs.S15, S16, and S17 provide computational results based on density function theory (DFT) at the *HSE06* level. These calculations confirm the metallicity of MXene with various terminal groups (-O, -F, and -OH) (Fig.S15). Anatase $TiO_2$, characterized by a suitable Fermi surface, bandgap, and work function (Fig.S16a-c), serves as the basis for establishing the semiconductor-metal Schottky interface with various MXene. Effective electron tunneling, influenced by the height and width of the Schottky barrier, enables a substantial number of carrier charges to traverse the Schottky barrier in various OBXene configurations (Fig.S16d-f)(*37*). Fig.S17a&b metaphorically illustrate how the interfacial Schottky barrier facilitates the directional migration of electrons. The charge redistribution and lattice distortion (Fig.S17c&d) result in a potential energy discontinuity and an interfacial dipole, accompanied by strong charge interactions at the interface. Meanwhile, charges effectuate out-of-plane transfer from $TiO_2$ to the MXene surface through Ti-O bonding, as shown in Fig.S17e, underscoring the role of orbit symmetry breaking in enhancing interfacial charge transfer.

The $TiO_2$ decorated within OBXene displays a distinct piezoelectric response under an electrical field, in contrast to the typical nonpiezoelectric nature of anatase $TiO_2$, attributed to the Schottky-induced asymmetric orbital configuration. Piezoresponse force microscopy (PFM) measurements of OBXene showed a linear increase in piezoelectric potential in conjunction with the deformation (Fig.S18a-c). In contrast, the Au-coated substrate displayed no such piezoelectric linearity, confirming that intrinsic piezoelectricity is not induced by metallic substrates such as monolayer MXene and Au. Employing dual-frequency resonance tracking (DFRT)-PFM to eliminate morphological interference, we obtained uniform and stable out-of-plane and in-plane piezoelectric signals from OBXene at an excitation voltage of 2.5 V (**Fig.5F**). Phase signals (**Fig.5G**) indicate consistent polarization orientations aligned with the applied electric field, indicating the presence of a piezoelectric dipole. The measured resonance peak intensities display a strong linear relationship with the excitation voltage (0–5 V) (**Fig.5H**), affirming its well-defined piezoelectric properties. The resonance peaks of OBXene are located at approximately 400 kHz and 704 kHz for out-of-plane and in-plane responses, respectively (**Fig.5I** and Fig.S19b), corresponding to the first and second harmonics(38), as shown below **Eq. (1)**:

$$\varepsilon = 2\,Q[\chi^2(E_0^2 + E_a^2 e^{i2wt} + 2E_0 E_a e^{iwt}) + P_s^2 + 2\chi(E_0 + E_a e^{iwt})P_s] \quad (1)$$

where $\varepsilon$, $Q$, $P_s$, $\chi$, and $w$ represent the electrostrictive strain, electrostrictive coefficient, spontaneous polarization, dielectric susceptibility, and angular frequency, respectively. Applied DC electric field, $E = E_0 + E_a e^{iwt}$, in which $E_0$ and $E_a$ represent the static component and time-varying component of the electric field, respectively.

Typical PFM experiments allow reasonable neglection of the static strain to focus on a simplified harmonic oscillator (SHO), as shown in **Eq. (2)**:

$$\varepsilon = 2\,Q\chi E_a(\chi E_0 + P_s)e^{iwt} + Q\chi^2 E a^2 e^{i2wt} \quad (2)$$

As predicted by the vertical and lateral response intensities, OBXene shows a much higher second-harmonic response than the first-harmonic response. This finding suggests that the linear excitation is not attributed to spontaneous polarization or intrinsic piezoelectricity in anatase $TiO_2$ but originates from the dipoles induced by the Schottky interface. However, the measured peak intensities on the Au substrate exhibited no linear dependence on the excitation voltage (Fig.S19c), indicating that the piezoelectric resonance signal in OBXene is not assigned to electrostatic effects. **Fig.5J** left shows a three-fold rotational symmetry plot of the second harmonic generation (SHG) intensity from a monolayer OBXene flake as a function of the crystal's azimuthal angle. The log−log plot in **Fig.5J** exhibits that the emission signal intensity is proportional to the square of the excitation laser power, confirming it is an SHG response. The appearance of a strong SHG signal indicates the inversion symmetry breaking in monolayer OBXene, which is attributed to Schottky-induced piezoelectric effect. The strong charge transfer among Schottky barrier, creating asymmetric electric dipoles, will further enhance the SHG response.

Therefore, we conclude that Schottky interface-induced band bending, combined with the built-in electric field, disrupts the inversion symmetry, which induces a polar structure and leads to the observed interface polar symmetry-induced piezoelectricity.

The piezoelectric Schottky junction in OBXene optimizes interlayered electron transport by regulating the bandgap, electrostatic potential, and density of states near the Fermi surface. Figs.S20, S21, and S22 illustrate the modulation of the Schottky barrier height and bandgap in anatase $TiO_2$ by external stress, considering various terminal groups (-F, -OH, -O) of $Ti_3C_2T_x$. Dynamic stress-induced lattice changes (positive and negative strains) are depicted in Fig.S23a-c, highlighting the role of orbit symmetry breaking in OBXene, especially with different terminal groups, in enhancing electron-hole pair separation under biaxial strain. Notably, the plotted projected density of state (PDOS) intensities in Fig.S23d-i, for both O and Ti atoms in anatase $TiO_2$ show significant enhancements under strain. In essence, the Ti-O bonding in the Schottky interface acts as a bridge for directional charge channels facilitated by orbit symmetry breaking. The uniform realignment of OBXene flakes in conductive ionogels establishes an 'energetic' and enables the utilization of the piezoelectric Schottky barrier to realize interlayered electron transport. Consequently, OBXene-filled ionogels achieve an ultrasensitive pressure response, as illustrated in **Fig.5K** and Fig.S24. Furthermore, Fig.S25 investigates additional properties of OBXene-filled ionogels, including self-healing capabilities, temperature sensitivity, and stretchability (Supplementary video S1-S3), which are retained from the pristine ionogel. These findings hold the potential to stimulate the development of a wide array of functional devices utilizing OBXene-filled ionogels. Fig.S25b presents the circuit diagram, which facilitates signal filtering and impedance-voltage conversion, to include the capacitive effect in the sensing response.

Our pressure mapping array features a multilayer design with the OBXene-filled ionogel in the middle and OBXene circuitries on the top and bottom to leverage its mechanical and electrical properties for high spatiotemporal recording (**Fig.5L**). Each sensing pixel is individually connected to OBXene-based traces on SEBS substrates, minimizing crosstalk between each sensing pixel (Fig.S26a)(*39*). Medium OBXene-filled ionogels, benefiting from the Schottky-induced piezoelectric effect, can realize efficient interlayered electron transport (**Fig.5M**). **Fig.5N-P** illustrate the experimental and corresponding finite-element-analysis (FEA) simulation to showcase the device's performance. Here, three irregular objects in the shape of letters, UNC, (weight, 11.4 g, 13.6 g, and 10.5 g, respectively), placed on the matrix generate a nonuniform distribution of pressure with additional variations in spatial tilt directions. The pressure mapping array, featuring 16 sensing pixels, continuously and synchronously captures the pressure distribution from the objects placed on top (Fig.S27). The color contrast mapping based on numerical simulation (**Fig.5O**) closely matches the measurement obtained through the mapping array (**Fig.5P**). Fig.S26b-d show additional measurements from a 2 x 2 mapping array in response to single-letter objects, showcasing its ability to accurately capture the pressure distribution. Additionally, Supplementary video S4 demonstrates the capability of this system to map dynamical variations in pressure distribution with 16-lead collecting signals in parallel, further confirming its utilization in real-time mapping of cardiac mechanics. Fig.S26e illustrates that OBXene-based arrays could distinctly identify different shapes and yield divergent stress distribution, emphasizing the high sensitivity of the OBXene-filling ionogels.

*In-vivo* **operation in porcine MI model**

Demonstrations of the OBXene patch with a porcine MI model provide comprehensive insights into its utility in clinical relevance with human cardiac physiology, including heart rate, cardiac action potential, and pathophysiologic effects of cardiac diseases. An OBXene patch is deployed on the left ventricle of a porcine heart (**Fig.6A**). *In-vivo* implantation of the OBXene patch integrates strain sensing, ECG sensing, thermal sensing, pressure mapping, thermotherapy, and electrotherapy (**Fig.6B**). The soft mechanics of the OBXene patch with a thin, biocompatible adhesive coating (BioGlue, CryoLife) at the bioelectronic-tissue interface ensures its conformal engagement with the beating heart and induces no observable disruptions to cardiac functions.

ECG and pressure mapping can reveal the dynamic evolution of cardiac electrophysiology and contraction over a broad spatiotemporal landscape, thus generating critical clinical insights inaccessible by other medical imaging technologies, such as ultrasound, magnetic resonance imaging, and X-ray computed tomography. The design and *in-vivo* deployment of an OBXene patch features 4 x 4 recording sites to map epicardial physiology (**Fig.6C**). The OBXene-based ECG array consists of 0.5-mm-diameter mini-pillar electrodes arranged in a square pattern with 1.5 mm adjacent intervals, featuring two reference electrodes for side-by-side comparison of simultaneously acquired ECG. The measured ECG mapping signals of healthy and infarcted hearts, which, by comparison, highlight the electrophysiological anomaly induced by MI (**Fig.6D**). To delineate the disparities between healthy and MI states, **Fig.6E** presents their recordings in a superimposed manner with an enlarged view of three channels over a single cardiac cycle. It is noteworthy that the ST segment in the MI model, corresponding to the plateau phase of the action potential, exhibits both depression and elevation relative to the baseline in sinus rhythm. This ST segment deviation indicates acute myocardial ischemia resulting from MI. The transition from the ST segment to the T-wave remains smooth, indicating that the repolarization processes of contractile cells are not yet affected in the MI model(*25*). Moreover, the latter part of the P-wave, which represents the depolarization of the left atrium, demonstrates a deviation from sinus rhythm, suggesting the presence of ectopic lesions (i.e., MI) outside the sinus node, affecting the left atrium's depolarization. Simultaneously, the PR segment, starting at the onset of the P-wave and ending at the onset of the QRS complex, reflecting the time interval from atrial to ventricular depolarization, reveals a notable change in amplitude(*21*). This indicates that MI also impacts conduction impulses through the atrioventricular node. **Fig.6F** presents ECG mapping extracted from four R peaks, as marked by vertical lines in **Fig.6D**. Here, instantaneous ECG values can be interpolated across the matrix and normalized, with black dots denoting the electrode positions. The calculated differential mapping in **Fig.6G** illustrates the propagated spatial disparities in electrophysiological activity between the sinus and MI models at various time intervals. In addition to localized amplified ECG details, this electrode matrix can record long-term ECG activity (Fig.S28a&b). Impressively, long-term ECG recordings maintain high SNR quality, providing insights into overall heart rhythm changes during depolarization and repolarization cycles (Fig.S28c). The high fidelity and spatiotemporal resolution of ECG recordings, from overall trends to detailed patterns, underscore the low bioelectronic-tissue interface impedance, attributable to orbit symmetry breaking in OBXene.

Spatiotemporal mapping of cardiac contractility enables precise localization of areas with reduced contractile function, particularly in local myocardial ischemia resulting from MI. The OBXene-filled ionogel, as discussed in **Fig.5K**, facilitates high-fidelity and high SNR pressure sensing. To achieve high spatial resolution mapping and control of contractile propagation on the porcine epicardium, **Fig.6H** shows a customized design that considers the edge-to-edge distance between neighboring sensing pixels in the pressure array in relevance with the physiological space constant of cardiac tissue. Notably, the entire circuit of the matrix is composed of OBXene, including the electrode plates and interconnected traces, owing to its low interface impedance. Fig.S29a presents a sequential filtering approach to mitigate environmental artifacts and motion-induced signal oscillations.

The 3D stacked dynamic mapping (**Fig.6J**) visually demonstrates localized differences in myocardial contraction between sinus rhythm and MI (**Fig.6I**) over time. Here, the pressure distribution is visualized via both the color and the terrain. The MI condition, represented by successive spatially interpolated frames, reveals reduced systolic amplitude and rhythmic disturbances within the cardiac cycle. Myocardial cell death and subsequent scar tissue formation in the MI condition compromise the contractile capacity of the heart, leading to diminished systolic force generation. The expanded and dynamic mapping in Fig.S29b and Supplementary video S5 offers further insights into arrhythmia pathogenesis and can be synchronized with structural computed tomography images of the heart for precise arrhythmia focus localization. Hence, the results demonstrate that OBXene-based sensing arrays can enable precise guidance for therapeutic interventions by identifying dysfunctional tissue, such as the source of abnormal rhythms in MI-induced ventricular tachycardia. The asymmetric orbital configuration in OBXene induces the piezoelectric effect, enables the high SNR and spatiotemporal resolved cardiac contraction mapping in OBXene-filled ionogel matrices.

**Wireless, battery-free operation in porcine model**

Postoperative interventions in MI are pivotal for optimizing patient outcomes, enhancing cardiac function, and preventing complications. Passive electromagnetic resonance sensing (PEMRS)(*40*) and magnetic resonance wireless power transfer (MRWPT)(*41*) allow distributed networks of sensors and actuators to measure and manipulate physiological activity throughout the body to promise precise and adaptive bioelectronic therapies with minimal risks or interference with daily activities(*42*). Here, we present a wireless, battery-free OBXene patch as a cardiac implant (**Fig.7A**). This patch harnesses the combined advantages of OBXene's symmetry-breaking properties along with enhanced power transfer and sensing capabilities provided by PEMRS and MRWPT. This enables real-time mechanical monitoring and closed-loop controlled electrical therapy for postoperative MI care. The pressure sensor employs two electrodes attached to both ends of a medium ionogel (dimensions: 3 cm x 1 cm x 2 mm, L x W x H, with a 2 wt. % filler) to form a capacitive pressure sensor that captures myocardial mechanics. The electrical pacing circuit, featuring OBXene electrodes, is printed on the bottom SEBS substrate and adhered to the porcine LV using BioGlue, as depicted in **Fig.7B** and Fig.S30a. Transmitting and receiving coils are subcutaneously implanted under the chest wall (Fig.S30b) and fully encapsulated with parylene C to ensure long-term biocompatibility and prolonged device longevity, demonstrating minimal local inflammatory responses and fibrosis in surrounding tissues within two weeks (Fig.S31).

Wireless sensing, based on the principles of PEMRS, relies on a passive inductor-capacitor (LC) resonance network formed by a soft inductor coil and capacitive pressure sensor, as illustrated in **Fig.7C** and Fig.S30c. During cardiac systoles and diastoles, the myocardium undergoes periodic changes in its surface pressure, reflected by the capacitive change of the OBXene-infused pressure sensor. This change can be quantified by measuring the shift in resonance frequency ($f_s$) of the LC network according to the equation:

$$f_s = \frac{1}{2}\pi\sqrt{LC} \quad (3)$$

where L and C are the inductance and capacitance of the resonance circuit, respectively. An external probe coil with minimal self-inductance is coupled with the implanted inductor coil to enable wireless readout of strain changes via a Vector Network Analyzer (VNA), measuring the shift in frequency at which the real impedance of the probe coil maximizes, achieving real-time, wireless, and fully passive measurement of cardiac mechanics. The wireless electrotherapy relies on MRWPT, which consists of an external power transmitter and an implanted soft receiver to harvest energy from the transmitter and convert it into electrical stimulation (**Fig.7D** and Fig.S30d).

**Fig.7E** and Supplementary video S6 provide an *in-vitro* demonstration of a wireless OBXene-based pressure sensor indicating a strong monotonic correlation between the measured resonant frequency and the pressure applied to the sensor. **Fig.7F** confirms the device's reliability one week after surgery, providing real-time data on myocardial contractility and monitoring the systolic interval (Supplementary video S7). In the MI porcine model, necrosis of myocardial tissue leads to a decrease in the number of viable myocytes that can actively contract, reducing the overall contractile force of the heart, while the necrotic tissue is replaced by noncontractile scar tissue (fibrosis), further disordering the myocardial contraction rhythm. Monitoring the process of cardiac tissue remodeling after an episode of MI again highlights the advantages of a long-term implantable wireless and battery-free device. Thus, wireless, battery-free OBXene-based devices hold promise in implementing long-term noninvasive post-MI prognostic assessment.

To further establish the therapeutic efficacy of the wireless OBXene patch, we conducted an in-situ experiment using an isolated porcine heart, as demonstrated in Fig.S30e. In this experiment, a PDMS pad (thickness 1.5 cm) separates the OBXene patch and the transmitter, simulating the power transmission barrier created by porcine skin. **Fig.7G** shows the bioelectrical signal resulting from cyclical stimulation from neighboring wireless pacing recorded by a 16-channel, OBXene-based array, with all 16 channels presenting conformal and time-synchronized output during cycles of pulsed electrical stimulation. The intercepted unit segments in **Fig.7H** reveal varying signal intensities in each channel due to differences in the spatial distribution of pacing propagation across the epicardium. Interpolated mapping (**Fig.7I**) captures spatiotemporal patterns of instantaneous pulses occurring during the "up" and "down" states, sorted from single-unit spikes. Three-dimensional finite element analysis (FEA, **Fig.7J**) simulates the distribution of the electric field during a DC 5 V input, with an electrical field approaching 2 V/mm near the inner electrode. This level of stimulation is sufficient to alleviate slow ventricular depolarization and conduction system dysfunction associated with acute myocardial ischemia(*43*), as discussed in **Fig.3D**. Effective propagation of electrical pacing signals across the epicardium enables synchronized ventricular and atrial contractions, which can be reflected in the wireless pressure sensor, facilitating closed-loop controlled postoperative cardiac pacing in MI.

**Discussion**

This study demonstrates the promising utility of orbit-symmetry-broken MXene (OBXene) in constructing soft bioelectronic implants, with advantages in interfacial impedance at the bioelectronic-tissue interface, amplified sensitivity, biocompatibility, manufacturing scalability, biostability, *in-vivo* longevity, and imaging visibility. Fundamentally, the symmetry-breaking architecture of OBXene focuses on coupling the out-of-plane charge transfer with the in-plane conduction to leverage charge transport across both inter- and intraflakes of such 2D material. This asymmetric configuration also induces the Schottky-induced piezoelectric effect, which amplifies the interlayered charge transport upon mechanical deformation. This characteristic of OBXene when used as a filler for conductive gels enables the construction of mechanical sensors with ultrahigh sensitivity and soft mechanics.

To translate this fundamental insight into application, we present an OBXene patch that integrates ECG mapping, pressure mapping, temperature sensing, strain sensing, thermotherapy, and electrotherapy on a single device platform that uses OBXene as the central functional material (e.g., sensor interfaces, interconnects, amplifier, transducer, and others). Both *in-vitro* and *in-vivo* experiments using Rat and Porcine MI models demonstrate the real-time, continuous monitoring of cardiac physiology and active thermo- and electrotherapy with closed-loop control capabilities. Furthermore, an *in-vivo* study with a living pig implanted with an OBXene patch for over one week demonstrates the device's wireless and battery-free sensing capabilities. This device has promising translational potential as a soft, multimodal, biocompatible implant. The current implantable sensing devices in the market may sense individual parameters, but few have the capability of sensing strain, ECG, and temperature in one multimodal electrode. In addition, there are no metal components in this OBXene-based cardiac patch. This implantable device will not cause artifacts that could negatively impact the quality of diagnostic images such as magnetic resonance imaging (MRI) or computed tomography (CT) images, making it ideal to integrate into the current healthcare diagnostic landscape.

The solution-processing compatibility for fabricating OBXene electrodes allows the construction of a high-density pixel array beyond the demonstrated 4 x 4 spatial resolution. The high charge mobility within the OBXene electrodes enables synchronous data transmission with more than 16 paths and shorter data transmission intervals to ensure overall data fidelity. The design of OBXene-based transistor arrays may offer an alternative to conventional multichannel arrays, highlighting potential future work. Furthermore, transistor arrays may enable integration with electronic circuits, such as amplification or filtering circuits, equipping soft implantable bioelectronic systems with on-chip or in situ signal processing. The major implications of OBXene described in this work, such as out-of-plane charge transport and Schottky-induced piezoelectricity, may also lead to expansive inventions of next-generation implantable electronics for various organs (brain, lung, kidney, etc.) and inspire the practical utility of broader low-dimensional materials in bioelectronic implants.

**Acknowledgment:** This work was supported by the start-up funds from University of North Carolina at Chapel Hill and the fund from National Science Foundation (award #

ECCS-2139659). Research reported in this publication was also supported by the National Institute of Biomedical Imaging and Bioengineering at the National Institutes of Health under award number 1R01EB034332-01.

**Author contributions:** Yizhang Wu, Yuan Li, and Yihan Liu equally contribute to this work. Y.W. and W.B. conceived the ideas and designed the research. Y.W., Y.L., S.X., S.L., B.D., L.Z., Z.Y., A.Z., W.X., Y.W., M.D., and K.H. fabricated and characterized the samples. Y.L., Y.W., and S.X. performed mechanical modelling and simulation. L.Y., D.Z., S.K., and Y.W. performed the ex-vivo and in-vivo studies. Y.L., S.X., N.L., H.W., M.W., G.Y., G.E., A.P., M.D., C.D., and H.K. performed the data analysis. Y.W., Y.L., K.H., and W.B. wrote the manuscript with input from all authors.

**Data and materials availability:** All data needed to evaluate the conclusions in the manuscript are present in the manuscript and/or the Supplementary Materials.

**Supplementary Materials**

**This PDF file includes:**

Supplementary Notes S1 to S16

Figs. S1 to S31

Legends for Movies S1 to S7

References

**Other supplementary material:**

Supplementary movie S1: Video of the response of OBXene-filled conductive ionogels to mechanical stretching.

Supplementary movie S2: Video of the response of OBXene-filled conductive ionogels to varying external temperatures.

Supplementary movie S3: Video of the self-healing capability of OBXene-filled conductive ionogels.

Supplementary movie S4: Video of the testing of an OBXene-filled ionogel matrix under pressurization conditions, accompanied by time-synchronized 16-lead responses collected via LabChart.

Supplementary movie S5: Video of continuous spatial mapping of myocardial contraction in real-time during both sinus rhythm and an MI model.

Supplementary movie S6: Video of the in-vitro mapping of the strain-stress-frequency relationship using a wireless strain sensor.

Supplementary movie S7: Video of real-time data acquisition two weeks after the implantation of a wireless stress sensor into a porcine heart.

Ghashghaee, N. Ghith, S. Giampaoli, S. A. Gilani, P. S. Gill, R. F. Gillum, E. V. Glushkova, E. V. Gnedovskaya, M. Golechha, K. B. Gonfa, A. H. Goudarzian, A. C. Goulart, J. S. Guadamuz, A. Guha, Y. Guo, R. Gupta, V. Hachinski, N. Hafezi-Nejad, T. G. Haile, R. R. Hamadeh, S. Hamidi, G. J. Hankey, A. Hargono, R. K. Hartono, M. Hashemian, A. Hashi, S. Hassan, H. Y. Hassen, R. J. Havmoeller, S. I. Hay, K. Hayat, G. Heidari, C. Herteliu, R. Holla, M. Hosseini, M. Hosseinzadeh, M. Hostiuc, S. Hostiuc, M. Househ, J. Huang, A. Humayun, I. Iavicoli, C. U. Ibeneme, S. E. Ibitoye, O. S. Ilesanmi, I. M. Ilic, M. D. Ilic, U. Iqbal, S. S. N. Irvani, S. M. S. Islam, R. M. Islam, H. Iso, M. Iwagami, V. Jain, T. Javaheri, S. K. Jayapal, S. Jayaram, R. Jayawardena, P. Jeemon, R. P. Jha, J. B. Jonas, J. Jonnagaddala, F. Joukar, J. J. Jozwiak, M. Jürisson, A. Kabir, T. Kahlon, R. Kalani, R. Kalhor, A. Kamath, I. Kamel, H. Kandel, A. Kandel, A. Karch, A. S. Kasa, P. D. M. C. Katoto, G. A. Kayode, Y. S. Khader, M. Khammarnia, M. S. Khan, M. N. Khan, M. Khan, E. A. Khan, K. Khatab, G. M. A. Kibria, Y. J. Kim, G. R. Kim, R. W. Kimokoti, S. Kisa, A. Kisa, M. Kivimäki, D. Kolte, A. Koolivand, V. A. Korshunov, S. L. K. Laxminarayana, A. Koyanagi, K. Krishan, V. Krishnamoorthy, B. K. Defo, B. K. Bicer, V. Kulkarni, G. A. Kumar, N. Kumar, O. P. Kurmi, D. Kusuma, G. F. Kwan, C. la Vecchia, B. Lacey, T. Lallukka, Q. Lan, S. Lasrado, Z. S. Lassi, P. Lauriola, W. R. Lawrence, A. Laxmaiah, K. E. LeGrand, M. C. Li, B. Li, S. Li, S. S. Lim, L. L. Lim, H. Lin, Z. Lin, R. T. Lin, X. Liu, A. D. Lopez, S. Lorkowski, P. A. Lotufo, A. Lugo, K. M. Nirmal, F. Madotto, M. Mahmoudi, A. Majeed, R. Malekzadeh, A. A. Malik, A. A. Mamun, N. Manafi, M. A. Mansournia, L. G. Mantovani, S. Martini, M. R. Mathur, G. Mazzaglia, S. Mehata, M. M. Mehndiratta, T. Meier, R. G. Menezes, A. Meretoja, T. Mestrovic, B. Miazgowski, T. Miazgowski, I. M. Michalek, T. R. Miller, E. M. Mirrakhimov, H. Mirzaei, B. Moazen, M. Moghadaszadeh, Y. Mohammad, D. K. Mohammad, S. Mohammed, M. A. Mohammed, Y. Mokhayeri, M. Molokhia, A. A. Montasir, G. Moradi, R. Moradzadeh, P. Moraga, L. Morawska, I. M. Velásquez, J. Morze, S. Mubarik, W. Muruet, K. I. Musa, A. J. Nagarajan, M. Nalini, V. Nangia, A. A. Naqvi, S. N. Swamy, B. R. Nascimento, V. C. Nayak, J. Nazari, M. Nazarzadeh, R. I. Negoi, S. N. Kandel, H. L. T. Nguyen, M. R. Nixon, B. Norrving, J. J. Noubiap, B. E. Nouthe, C. Nowak, O. O. Odukoya, F. A. Ogbo, A. T. Olagunju, H. Orru, A. Ortiz, S. M. Ostroff, J. R. Padubidri, R. Palladino, A. Pana, S. Panda-Jonas, U. Parekh, E. C. Park, M. Parvizi, F. P. Kan, U. K. Patel, M. Pathak, R. Paudel, V. C. F. Pepito, A. Perianayagam, N. Perico, H. Q. Pham, T. Pilgrim, M. A. Piradov, F. Pishgar, V. Podder, R. V. Polibin, A. Pourshams, D. R. A. Pribadi, N. Rabiee, M. Rabiee, A. Radfar, A. Rafiei, F. Rahim, V. Rahimi-Movaghar, M. H. U. Rahman, M. A. Rahman, A. M. Rahmani, I. Rakovac, P. Ram, S. Ramalingam, J. Rana, P. Ranasinghe, S. J. Rao, P. Rathi, L. Rawal, W. F. Rawasia, R. Rawassizadeh, G. Remuzzi, A. M. N. Renzaho, A. Rezapour, S. M. Riahi, R. L. Roberts-Thomson, L. Roever, P. Rohloff, M. Romoli, G. Roshandel, G. M. Rwegerera, S. Saadatagah, M. M. Saber-Ayad, S. Sabour, S. Sacco, M. Sadeghi, S. S. Moghaddam, S. Safari, A. Sahebkar, S. Salehi, H. Salimzadeh, M. Samaei, A. M. Samy, I. S. Santos, M. M. Santric-Milicevic, N. Sarrafzadegan, A. Sarveazad, T. Sathish, M. Sawhney, M. Saylan, M. I. Schmidt, A. E. Schutte, S. Senthilkumaran, S. G. Sepanlou, F. Sha, S. Shahabi, I. Shahid, M. A. Shaikh, M. Shamali, M. Shamsizadeh, M. S. R. Shawon, A. Sheikh, M. Shigematsu, M. J. Shin, J. Il Shin, R. Shiri, I. Shiue, K. Shuval, S. Siabani, T. J. Siddiqi, D. A. S. Silva, J. A. Singh, A. Singh, V. Y. Skryabin, A. A. Skryabina, A. Soheili, E. E. Spurlock, L. Stockfelt, S. Stortecky, S. Stranges, R. S. Abdulkader, H. Tadbiri, E. G. Tadesse, D. B. Tadesse, M. Tajdini, M. Tariqujjaman, B. F. Teklehaimanot, M. H. Temsah, A. K. Tesema, B. Thakur, K. R. Thankappan, R. Thapar, A. G. Thrift, B. Timalsina, M. Tonelli, M. Touvier, M. R. Tovani-Palone, A. Tripathi, J. P. Tripathy, T. C. Truelsen, G. M. Tsegay, G. W. Tsegaye, N. Tsilimparis, B. S. Tusa, S. Tyrovolas, K. K. Umapathi, B. Unim, B. Unnikrishnan, M. S. Usman, M. Vaduganathan, P. R. Valdez, T. J. Vasankari, D. Z. Velazquez, N.

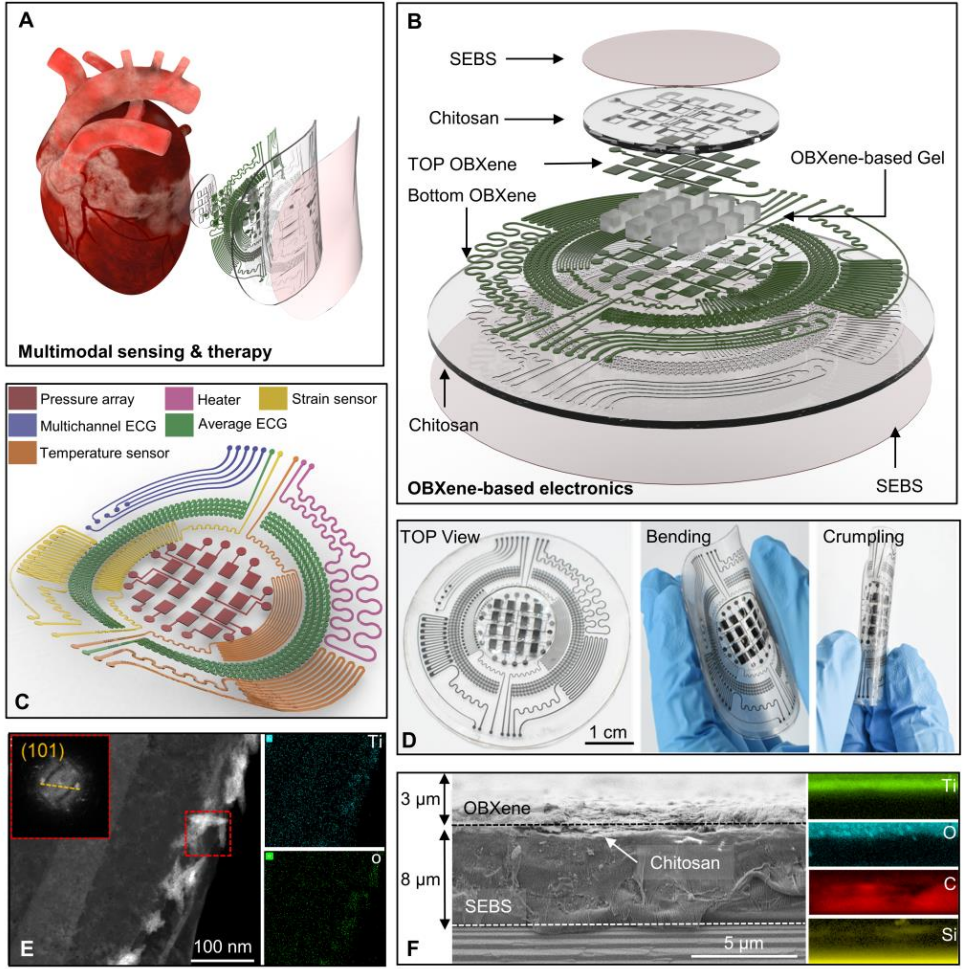

**Figure 1. Overview of the OBXene patch for Multimodal Sensing and Electrotherapy. A.** Schematic illustration of a soft, deformable OBXene patch attached to an epicardial surface for multimodal sensing-guided autonomous therapy. **B.** Schematic illustration with an exploded view of the OBXene patch. The constituent components include pliable styrene-ethylene-butylene-styrene (SEBS) substrates, chitosan that serves as a cross-linking agent, upper and lower OBXene traces, and a 4x4 active matrix embedded with an OBXene-filled ionogel (poly(acrylic acid)-copolymer-poly(acrylamide), labeled P(AA-co-AAm)). **C.** Schematic illustration highlighting individual functional units that are integrated on a single OBXene patch platform. **D.** Optical image displaying the mechanical robustness of an OBXene patch under various deformations. **E.** Image of high-angle annular dark-field imaging-scanning transmission electron microscopy (HAADF-STEM) illustrating the fusiform anatase $TiO_2$ anchored on the border of monolayered MXene. The attached image of electron energy loss spectroscopy (EELS) presents the elemental distribution of Ti and O in the highlighted region. **F.** Image of lateral scanning electron microscopy (SEM) depicting the cross-sectional view of the OBXene patch to elaborate on the elemental composition and distribution within the layered architecture.

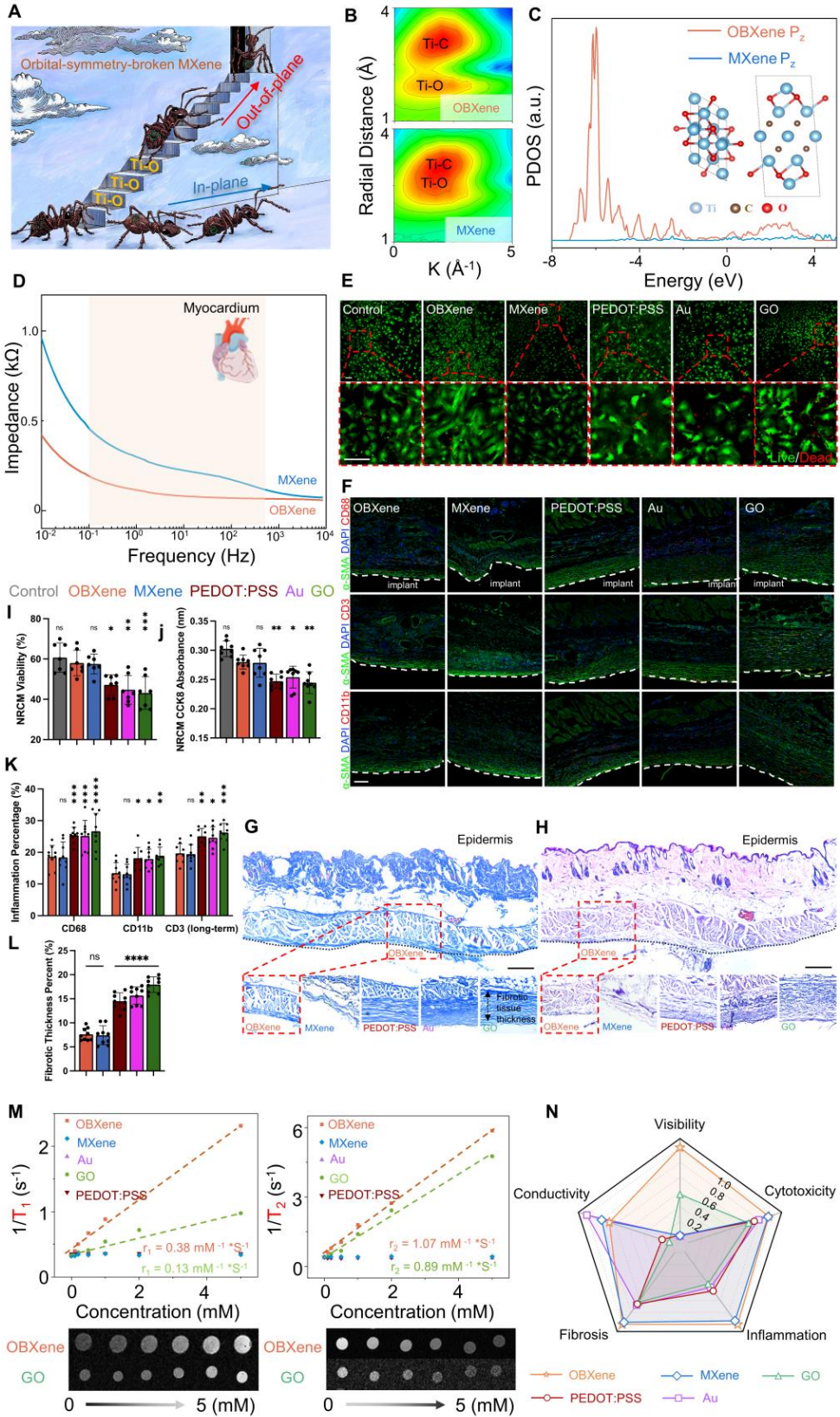

**Figure 2. Orbital symmetry breaking in OBXene optimizes the bioelectronic system. A.** Metaphorical illustration of OBXene, illustrating the activation of the out-of-plane electron transfer pathway via the marginal derived Ti-O bonding. The moving ants simile transferring electrons. The enabled isotropic electron-transferring freedom enhances charge mobility at the OBXene-biotic interface. **B.** Morlet wavelet transforms of Ti K-edge $k^2$-weighted EXAFS spectra for OBXene (top) in comparison to the lyophilized monolayered MXene (bottom). EXAFS, extended X-ray absorption fine structure spectroscopy. **C.** PDOS of O atoms projected on the $P_z$ orbital in OBXene and MXene. PDOS, projected density of states. **D.** Electrochemical impedance spectra (EIS) measured in 5 mM $K_4[Fe(CN)_6]/K_3[Fe(CN)_6]$ (1:1) solution containing KCl (0.1 M) for the OBXene electrode in comparison to MXene electrodes. The highlighted regions mark the inherent bioelectric frequency range of myocardium. **E.** Cytotoxicity analysis of Control, OBXene, MXene, PEDOT:PSS, Au, and GO implants (Dimension, 2.5 mm x 2.5 mm x 50 μm, L x W x H) in neonatal rat cardiomyocytes (NRCMs), indicating higher cardiomyocyte viability in the OBXene group. Green indicates live cells, while red indicates dead cells. **F.** Immunohistochemical analysis of tissues surrounding the implanted OBXene, MXene, PEDOT:PSS, Au, and GO patches. The tissues were stained with α-smooth muscle actin (α-SMA) (smooth muscle and myofibroblast), DAPI (nucleus), and inflammatory markers: CD68 (macrophages), CD3 (T cells), and CD11b (neutrophils). **G, H.** Microscopy images of the surrounding tissues with (**G**) Masson's trichrome staining indicating less fibrotic tissue thickness in the OBXene group and (**H**) hematoxylin and eosin (H&E) staining indicating fewer inflammatory cells in the OBXene group. **I.** Quantification of NRCM viability treated with the control group and all five experimental groups (with OBXene, MXene, PEDOT:PSS, Au, and GO). OBXene did not show a significant viability decrease compared to the control. **J.** Cell Counting Kit 8 (CCK8) analysis with NRCM showed less cardiomyocyte viability attenuation in the OBXene group than in the other experimental groups. **K.** Quantification of inflammatory marker expression in rat skin indicating that the OBXene patch induced fewer macrophages (CD68), T cells (CD3), and neutrophils (CD11b) than patches of MXene, PEDOT:PSS, Au, and GO, respectively. **L.** Quantification of rat skin fibrotic tissue thickness percent indicating that OBXene induced less fibrotic tissue compared to patches of MXene, PEDOT:PSS, Au, and GO. Scale: **E, F** 100 μm; **G, H** 400 μm. All data are means ± SDs. Group comparisons were performed using **I-L** one-way and **K** two-way ANOVA followed by post hoc Bonferroni correction. Significance: ns indicates $P > 0.5$, * $P \leq 0.05$, ** $P \leq 0.01$, *** $P \leq 0.001$, **** $P \leq 0.0001$. **M.** 7.0T-MRI (magnetic resonance imaging) linear relationship between $T_1$-weighting (left)/$T_2$-weighting (right) and different concentrations of OBXene, MXene, Au, GO, and PEDOT:PSS. (Below: *in-vitro* MRI images of OBXene and GO at different concentrations). **N.** Radar charts qualitatively summarizing the performance of OBXene as a bioelectronic system compared with other common bioelectronic materials. Please refer to the supplementary material for the quantitative criteria for the different indicators of the radar chart.

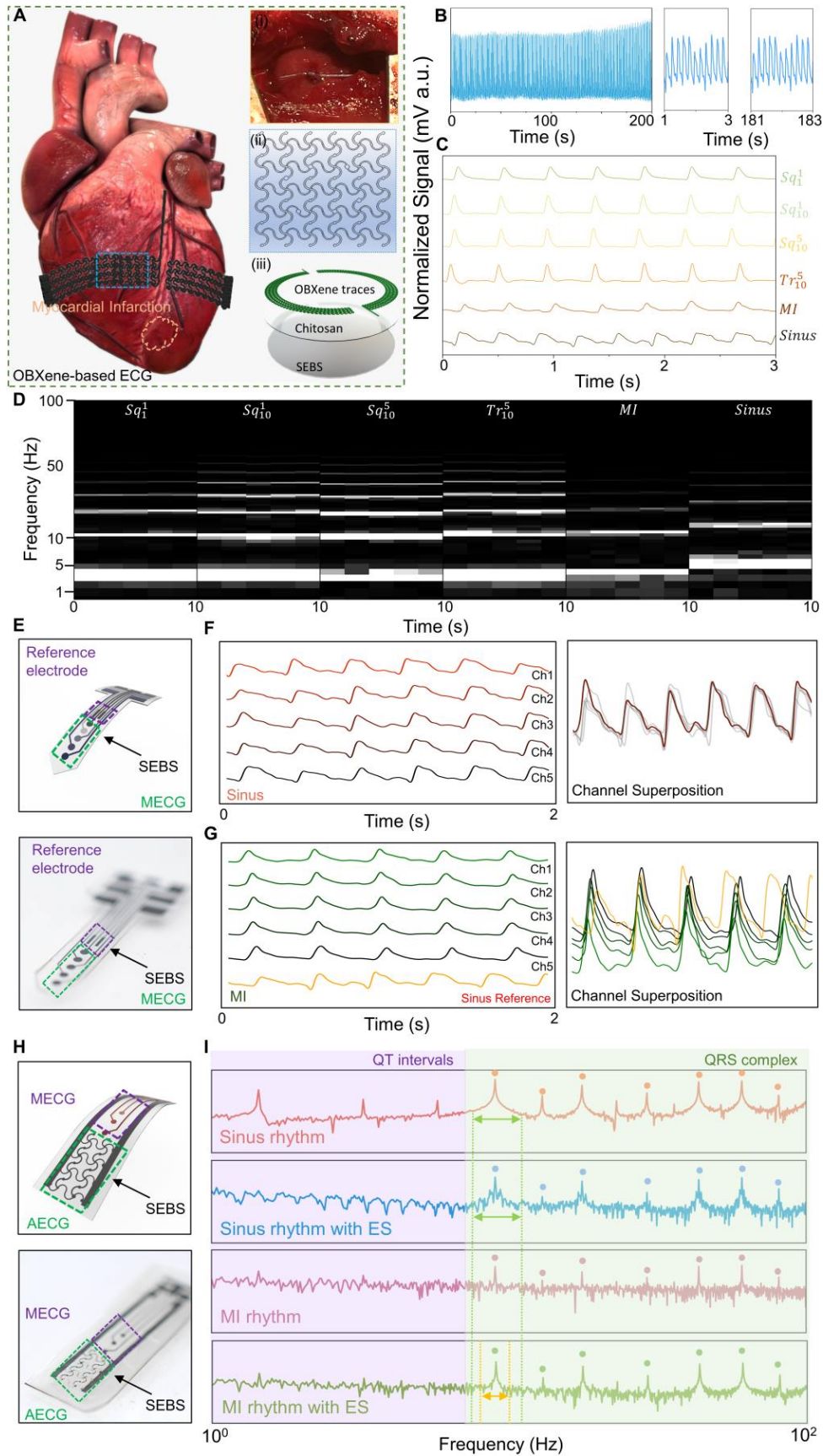

**Figure 3. ECG electrodes based on OBXene traces exhibiting sensitive and stable recording of epicardial electrophysiological activities. A.** Schematic illustration of the electrocardiograph (ECG) traces mounted on the rat epicardial surface with an infarcted coronary artery. Insets: (i) Optical image of the coronary artery ligation. (ii) Magnified image highlighting the structural design of the AECG traces based on a space-filling serpentine geometry. (iii) Spatial setup of the OBXene-based average ECG (AECG) traces. **B.** Left: Representative AECG recording with the OBXene traces on the beating heart of a rat. Right: Zoomed-in view of the collected AECG signals intercepted from the early and terminal stages. **C.** Measured AECG signals of a rat heart at a healthy stage (sinus), with myocardial infarction (MI), and undergoing various electrotherapy. The sinus rhythm and MI rhythm without electric stimulus are displayed as references at the bottom (black line and brown line, respectively). The plots above the reference demonstrate the measurement under myocardial infarction with various electrotherapy patterns. From top to bottom: a square wave with 1 V amplitude and 1 Hz frequency ($Sq_1^1$), a square wave with 1 V amplitude and 10 Hz frequency ($Sq_{10}^1$), a square wave with 5 V amplitude and 10 Hz frequency ($Sq_{10}^5$), and a triangle wave with 5 V amplitude and 10 Hz frequency ($Tr_{10}^5$). **D.** Frequency spectrogram of AECG recordings under various levels of electrical stimulation. **E.** Schematic view (top) and optical image (bottom) of the OBXene-based multichannel ECG (MECG) electrodes. **F, G.** Left: Representative MECG data of sinus rhythm (**F**) and myocardial infarction (**G**), recorded with the MECG electrodes. Right: The superposition of the five channels highlighting the abnormal electrophysiological activities resulting from MI. **H.** Schematic illustration (top) and optical image (bottom) of the combination of MECG and AECG traces. **I.** Frequency spectrogram of sinus and myocardial infarction rhythm under in situ autonomous electrotherapy, measured with MECG and AECG traces. The measured ECG intervals (QT interval and QRS complex) are consistent with the heart rate variability of rats. By comparison, the measured ECG signals of the MI heart with electrical stimulation (ES) show an observable enhancement in cardiac regulation.

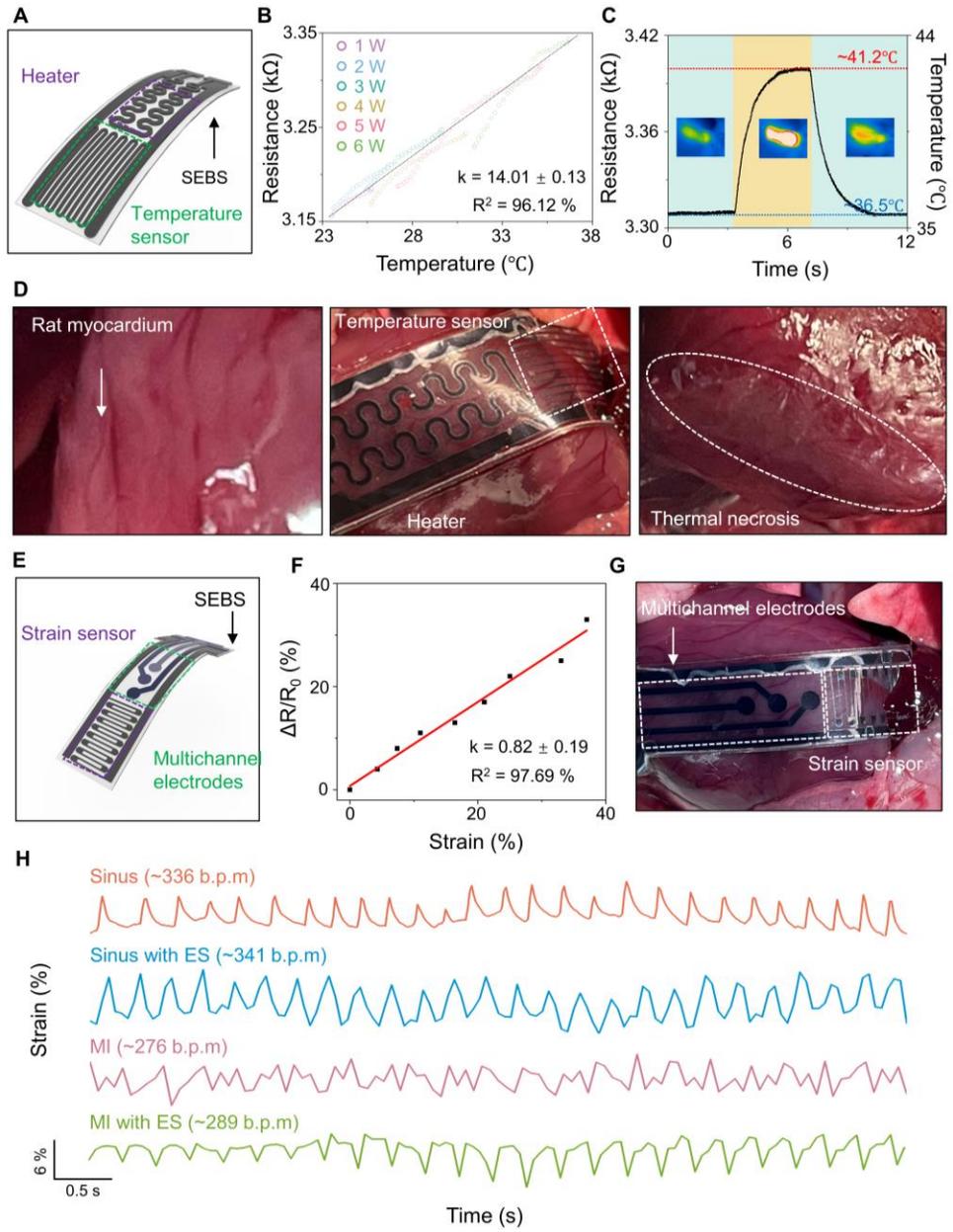

**Figure 4. OBXene-based bioelectronic device realizing multimodal sensing and feedback. A.** Schematic view of the combination of a temperature sensor and heater based on OBXene traces. **B.** The temperature-heater sensor achieves ideal linear regression under various powers. **C.** Real-time temperature-change sensing of the temperature-heater sensor before and after contact with the rat myocardium. Insets show the thermal phase diagrams before (left), during (middle), and after (right) applying a voltage of 5 W to the heater. **D.** The corresponding optical image of the temperature-heater sensor mounted on rat epicardium and the thermal ablation using the OBXene-based heater. **E.** Schematic view of the combination of multichannel electrodes and strain sensors based on OBXene traces. **F.** Calibration curve of the OBXene-based strain sensor under various strains. **G.** Optical image of the multichannel electrode-strain sensor mounted on rat epicardium. **H.** The corresponding measurements of the multichannel electrodes-strain sensor on a rat myocardium during sinus and MI rhythm, with the in-situ electrotherapy attached. The contraction of the strain sensor estimates bpm in each case.

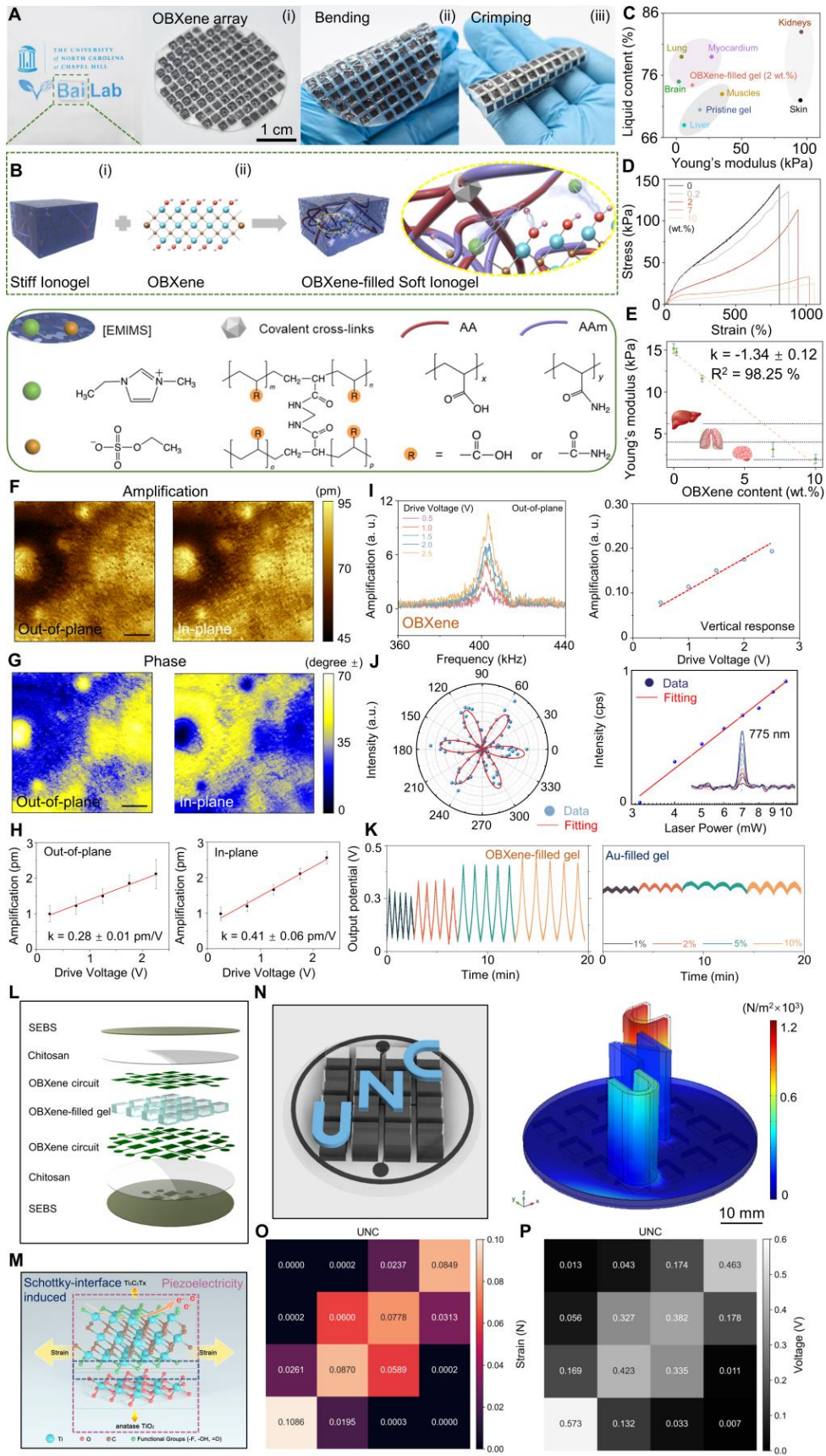

**Figure 5. OBXene-filled ionogel facilitates spatially resolved pressure mapping. A.** (i) Image of a conductive P(AA-co-AAm) ionogel filled with OBXene. (ii) Image showing a 124-unit OBXene-filled ionogel array incorporating a flexible SEBS substrate and supportive OBXene pad, which collectively serve as the core sensing interface for pressure mapping. (iii) Images demonstrating the flexibility and structural integrity of the array under various mechanical deformations. **B.** Schematic illustration showing a molecular network of the OBXene-filled conductive ionogel. The preparation process is described in two sections: (i) 1-ethyl-3-methylimidazolium ethyl sulfate (EMIES) coupling with poly(acrylic acid) (PAA) polymers and Poly (acrylamide) (PAAm) polymers to form a stiff but brittle polymer network. (ii) The electronegative OBXene flakes are uniformly distributed in the matrix of the ionogel via hydrogen bonding with an electrostatic self-assembly mechanism. The ingredients of the ionogel and corresponding molecular structure formula are listed below. **C.** Comparison of Young's modulus and liquid content among pristine ionogel, conductive ionogel with 2 wt.% OBXene filler, and some representative organs and tissues for reference. **D**. Measured tensile stress–strain curves of ionogels with various filling amounts. Each sample is stretched until failure. **E.** Young's modulus of gels with various concentrations of fillers, indicating a linear relation between Young's modulus and the OBXene ratio. The measured range of Young's modulus matches those of various tissues. Referenced Young's moduli of the liver, lung, and brain are provided by the IT'IS Foundation. **F.** DFRT-PFM scanning for out-of-plane and in-plane amplitude located on OBXene at an excitation voltage of 2.5 V. Voltage interval is 0.5 V. DFRT-PFM, dual-frequency resonance tracking-piezoresponse force microscopy. The scale bar is 100 nm. **G.** Corresponding out-of-plane and in-plane phase signals show the presence of a piezoelectric dipole on the surface. The orientation of the electrical field is perpendicular to the plane. **H.** Out-of-plane (left) and in-plane (right) statistical amplitudes on OBXene concerning the drive voltage, both illustrating a clear linear relationship. **I.** Vertical resonance peaks (left) of OBXene at approximately 412 kHz with gradient voltages, showing linearity of the vertical resonance peak (right) intensity concerning the drive voltage. **J.** Polar plot (left) and excitation power plot (right, log-log scale) of the SH intensity from monolayer OBXene as a function of the crystal's azimuthal angle θ. Inset: SHG spectra of the different excitation powers. Angle dependence of SHG intensity and for monolayer OBXene, while the SH component measured perpendicular to the excitation polarization. **K.** The real-time relative voltage output of OBXene-filled gel (left) and Au-filled gel (right) is consistent with cyclic strains, indicating a significantly higher responsivity from OBXene-filled gel under the same strain. **L.** Schematic illustration with an exploded view of an OBXene-filled pressure mapping array with 4x4 units, containing the top and bottom circuits based on OBXene traces and OBXene-filled ionogels as the medium. **M.** Schematic illustration of OBXene-filled gel achieving sensitive and stable pressure feedback ascribed to the piezoelectricity stemming from Schottky interface-induced inversion symmetry breaking. **N.** Schematic illustration (left) of 3D-printed objects (emulating resinous letters, U, N, and C, respectively) placed on the pressure array. **O.** The corresponding numerical distribution of simulation of the array. **P.** Distributive voltage signals measured by the pressure mapping array. The data are plotted as the means with n = 10 per unit corresponding to SDs (standard deviations). Data is gathered by the signal conditioning circuit of the impedance conversion circuitry.

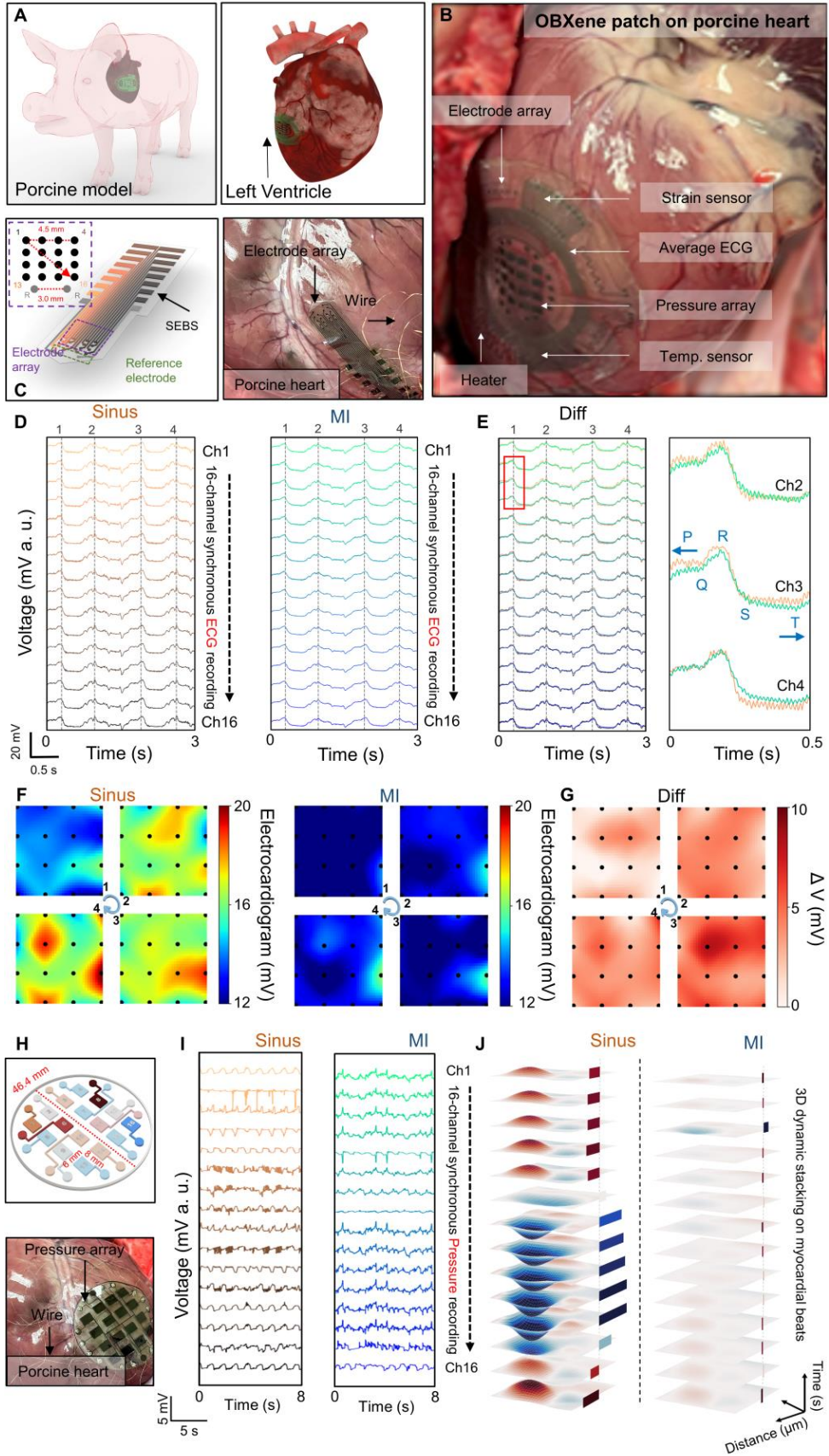

**Figure 6. Physiological signal mapping in porcine model recorded by OBXene patches. A.** Schematic illustrations of (left) an OBXene patch deployed onto a porcine heart and (right) placement of the OBXene patch onto the porcine heart near the left ventricle. **B.** Optical image showing an OBXene patch deployed onto a porcine heart. The OBXene patch includes an active pressure matrix and multifunctional sensing modules. **C.** Schematic illustrations of a (left) 4 x 4 OBXene-based electrode array. The inset shows a map exhibiting the dimensional parameters of the design, and (right) the electrode array conformally attached to a porcine heart. **D.** Segments of the recorded EEG signal from the electrode array on a living porcine heart with sinus rhythm (left) and MI (right). The data are displayed following the spatial arrangement of the electrode array (**C**). **E.** Zoomed-in views showing the difference of a single ECG unit intercepted from three channels, marked with salient ECG features. **F.** Instantaneous electrophysiology mapping corresponds to sinus rhythm (left) and MI molding (right). The mapping frames selected here correspond to the respective four synchronous points highlighted by the vertical dashed lines in (**D**). **G.** Calculated differential mapping of electrophysiological activity by comparing measurements of the relative activation time segment between sinus and MI. **H.** Top: Schematic illustration of an OBXene patch with a 4 x 4 pressure sensing array, illustrating the size and serial number of each unit. Bottom: Optical image of the OBXene patch conformally attached to a porcine heart. **I.** Representative real-time synchronous recording of the OBXene patch through highpass, lowpass, and bandstop filtering on a porcine heart in the sinus state (Left) and MI molding state (Right). The configuration of each channel is consistent with (**H**). **J.** The corresponding 3D dynamic mappings (sinus state (Left) and MI molding state (Right)) on the myocardial contract are stacked by interpolation and normalization. Specifically, a red and peak terrain indicates high pressure, and a blue and valley terrain indicates low pressure. Furthermore, the sidebar demonstrates the average pressure of each frame. A red bar exhibits an average pressure higher than the baseline, while vice versa signifies a lower average pressure. The X-Y plane illustrates the spatial distribution of each sensing unit. Vertical orientation depicts the temporal flow, with a 40 ms time interval for each frame.

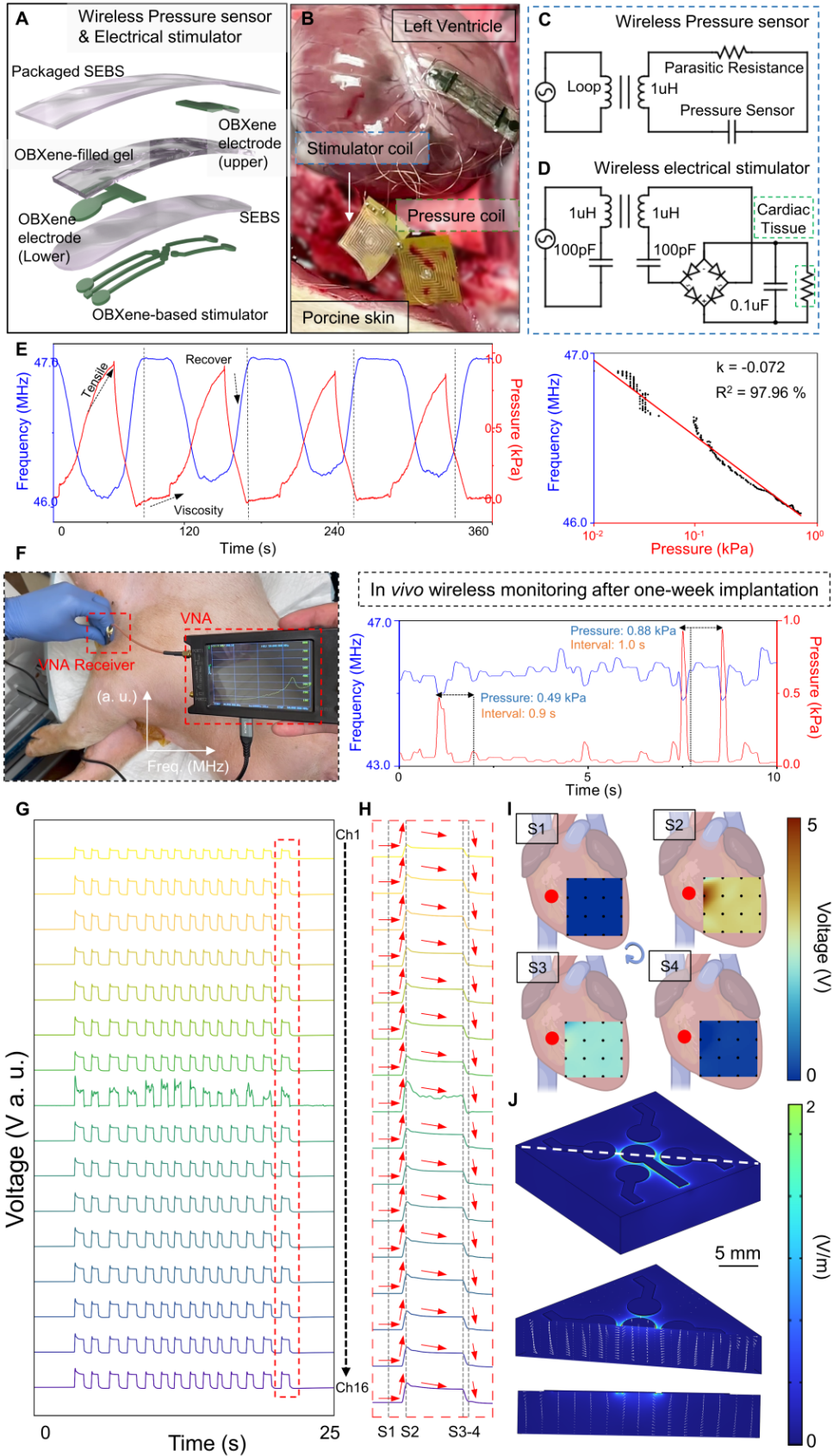

**Figure 7. Real-time and closed-loop wireless OBXene patch demonstrating simultaneous colocalized monitoring and therapy in a porcine model. A.** An exploded schematic illustration of a wireless OBXene patch integrated with a pressure sensor and an electrical stimulator. The upper-pressure sensor is composed of two electrodes based on OBXene traces, OBXene-filled gels, and the SEBS encapsulation layer. The electrical stimulator consists of four surrounding electrodes and a central reference, all based on OBXene traces, encapsulated by parylene C via shadow masks. **B.** Optical image showing the wireless OBXene patch mounted on a porcine heart. The stimulator coil and strain coil were both placed subcutaneously with a soft, biocompatible wire for optimized signal and energy transmission. **C.** Equivalent circuit diagram of the component for wireless pressure sensing. Connecting the OBXene-filled capacitor to an inductor coil forms an LC resonance circuit, where the pressure change on the capacitor leads to the capacitive change and translates to the change in the characteristic resonance frequency of the LC circuit, which can be captured by an external probing coil connected to a vector network analyzer (VNA) to realize the wireless sensing function. **D.** Equivalent circuit diagram of the component for wireless stimulation. **E.** Cyclic time-synchronized stretch curves of ionogels *in vitro* (Left), establishing the linear mapping among frequency-pressure upon 25 mm x 10 mm x 2 mm (L x W x H) medium (Right). The maximum strain is up to 30 %. **F.** Wireless and real-time detection of porcine cardiac contractility and heartbeat intervals after one week of implantation. **G.** Sixteen-channel electrode array based on OBXene traces shows the stimulation received from a wireless electrical stimulator. **H.** Single-unit spikes sorted from the data. **I.** Instantaneous voltage mapping recorded across the 4 x 4 OBXene-based electrode array during the before state (panel S1), up state (panel S2), down state (panel S3), and after state (panel S4). The timing of these voltage maps is indicated in (**H**) by the vertical lines. Red dots and interpolated plots showing the spatial position of the electrical stimulator and 16-lead electrodes on the porcine heart. **J.** Electrostatic Finite Element Analysis (FEA) results of the Electrical Field Distributions in the myocardium of the OBXene-based stimulation electrodes. Scale bar, 5 mm.